\documentclass[aps,prb,twocolumn,floatfix,showpacs,citeautoscript,longbibliography,hyperlinks]{revtex4-1}

\usepackage{graphicx}
\usepackage{dcolumn}
\usepackage{bm}
\usepackage{bbold}
\usepackage{mathtools}
\usepackage{color}
\usepackage{transparent}
\usepackage[colorlinks=true,citecolor=blue]{hyperref}

\begin{document}

\title{Lee-Yang zeros and large-deviation statistics of a molecular zipper}

\author{Aydin Deger}
\author{Kay Brandner}
\author{Christian Flindt}
\affiliation{Department of Applied Physics, Aalto University, 00076 Aalto, Finland}

\date{\today}

\begin{abstract}
The complex zeros of partition functions were originally investigated by Lee and Yang to explain the behavior of condensing gases. Since then, Lee-Yang zeros have become a powerful tool to describe phase transitions in interacting systems. Today, Lee-Yang zeros are no longer just a theoretical concept; they have been determined in recent experiments. In one approach, the Lee-Yang zeros are extracted from the high cumulants of thermodynamic observables at finite size. Here, we employ this method to investigate a phase transition in a molecular zipper. From the energy fluctuations in small zippers, we can predict the temperature at which a phase transition occurs in the thermodynamic limit. Even when the system does not undergo a sharp transition, the Lee-Yang zeros carry important information about the large-deviation statistics and its symmetry properties. Our work suggests an interesting duality between fluctuations in small systems and their phase behavior in the thermodynamic limit. These predictions may be tested in future experiments.
\end{abstract}

\maketitle

\section{Introduction}

Continuity is a central concept in physics. Often the properties of a system change only slightly in response to small variations of an external control parameter. Phase transitions, however, are a remarkable exception to this physical principle.\cite{Callen1985,Chandler1987,Goldenfeld1992} Consider for example a macroscopic gas in contact with a heat reservoir. If the temperature is slowly decreased, the gas may suddenly condense into a liquid and the density of molecules correspondingly increases abruptly. But how do the molecules know that they should suddenly form a liquid rather than a gas? Born and Fuchs first posed this question,\cite{born1938} which for a long time defied a convincing answer.

An early attempt to explain the gas--liquid transition was made by Maxwell and Van der Waals.\cite{maxwell1874} By modifying the ideal gas law to account for the attractive forces between the molecules, they were able to qualitatively explain several experimental observations. However, their theory relied on phenomenological arguments and did not explain how the microscopic configuration of the fluid changes at the condensation point. A more rigorous approach based on statistical mechanics was later proposed by Mayer and coworkers \cite{mayer1937,mayer1937a,Mayer1941}. They expanded the free energy of the gas in powers of its density so that the thermodynamic limit could be taken for each term in the expansion. Their cluster expansion provided a quite accurate description of the gas phase. However, it could not correctly account for the properties of the liquid phase.

\begin{figure}
	\centering
	\includegraphics[width=0.95\columnwidth]{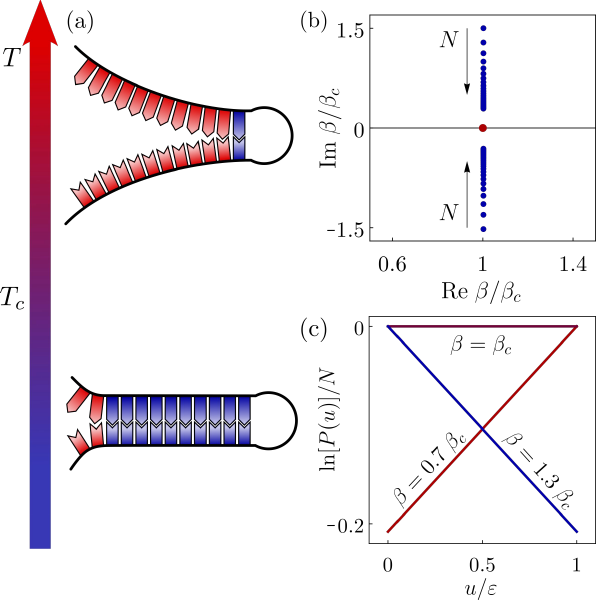}
	\caption{Zipper model, Lee-Yang zeros, and large-deviation statistics. (a) The molecular zipper is a double-stranded macromolecule held together by $N$ links that can be either open (energy $\varepsilon$) or closed (energy 0). At the transition temperature $T_c$, the system goes through a phase transition. (b) Lee-Yang zeros in the complex plane of the inverse temperature $\beta=1/(k_BT)$. (c) Large-deviation statistics of the energy per link for large $N$. Distributions are shown for three different temperatures, $\beta=0.7\beta_c$, $\beta_c$, $1.3\beta_c$, with $\beta_c=1/(k_BT_c)$.}
\label{fig1}
\end{figure}

This drawback of Mayer's method motivated Lee and Yang to develop a non-perturbatitive approach.\cite{Yang1952a,Lee1952} Instead of using an expansion, they analytically continued the partition function into the complex plane by allowing the control parameter to acquire an imaginary part. The zeros of the partition sum in the complex plane of the control parameter are the central objects of this framework. For small volumes, the zeros are well separated from the real axis. As the volume increases, the number of Lee-Yang zeros grows and they move towards the phase transition point on the real-axis which they reach in the thermodynamic limit. Since the zeros of the partition function correspond to logarithmic singularities of the free energy, it becomes clear that Mayer's expansion could not converge beyond the critical point. Moreover, one may consider the zeros in the complex plane of any control parameter, e.~g.~fugacity,\cite{Yang1952a} temperature,\cite{fisher1965,grossmann1967,grossmann1969a,grossmann1969} or magnetic field,\cite{Lee1952} implying that the arguments of Lee and Yang are not restricted to the gas--liquid transition. Rather, they show that phase transitions generally cannot be described by conventional perturbation methods.

Over the last decades, the ideas of Lee and Yang have developed into a powerful theoretical tool in statistical physics\cite{Blythe2003,bena2005} with applications to diverse problems ranging from protein folding \cite{Lee2013a,Lee2013b} over percolation\cite{Arndt2001,Dammer2002} and complex networks\cite{Krasnytska2015,Krasnytska2016} to Bose-Einstein condensation.\cite{Borrmann2000,Mulken2001,Dijk2015,Gnatenko2017} Further extensions include the use of Lee-Yang zeros to characterize phase transitions in non-equilibrium systems,\cite{Blythe2002,Blythe2003,bena2005}
quenched quantum systems,\cite{Heyl2013,Azimi2016} and glass formers \cite{Merolle2005,Garrahan2007,Hedges2009,Speck2012}. In addition, a number of experiments have shown that Lee-Yang zeros are not just a theoretical concept; they can also be experimentally determined. In one experiment, the density of Lee-Yang zeros was obtained from magnetization data of a two-dimensional Ising ferromagnet.\cite{Binek1998} More recently, Lee-Yang zeros have been determined by measuring the quantum coherence of a probe spin coupled to an Ising-type spin bath.\cite{Wei2012,Peng2015} In the context of non-conventional phase transitions,\cite{Garrahan2010} Lee-Yang zeros have been determined from real-time measurements of the activity along a quantum-jump trajectory.\cite{flindt2013,Brandner2017} For finite-size systems, the Lee-Yang zeros can be extracted from the cumulants of the observable which is conjugate to the control parameter.\cite{flindt2013,Brandner2017}  With increasing system size, one can then extrapolate the position of the Lee-Yang zeros in the thermodynamic limit and predict a possible phase transition. In the experiment of Ref.~\onlinecite{Brandner2017}, the Lee-Yang zeros converged to points that were slightly off from the real-axis, corresponding to a crossover rather than a sharp phase transition. Still, the Lee-Yang zeros carried important information about the large-deviation statistics,\cite{touchette2009} which is typically difficult to measure.

The purpose of this paper is to show that the method developed in Refs.~\onlinecite{flindt2013,Brandner2017} is not restricted to the extraction of dynamical Lee-Yang zeros from quantum-jump trajectories. The method can in fact be employed to a large class of problems from statistical mechanics. As a specific example, we analyze a simple model\cite{Kittel1969} of a one-dimensional DNA molecule\cite{Poland1966,Kafri2000,Bar2007} as depicted in Fig.~\ref{fig1}~(a). The model describes a molecular zipper consisting of a double-stranded macromolecule with infinite-range interactions. We extract the Lee-Yang zeros in the complex plane of the inverse temperature from the energy cumulants of short zippers, Fig.~\ref{fig1}~(b). We can then accurately predict the temperature at which the system undergoes a phase transition in the thermodynamic limit. As we show, the Lee-Yang zeros carry important information about the large-deviation statistics of the energy as illustrated in Fig.~\ref{fig1}~(c). As an extension, we introduce a finite energy cost for breaking the zipper. In this case, the system exhibits a crossover rather than a sharp phase transition. The Lee-Yang zeros can still be extracted from the energy fluctuations in zippers of finite size and they determine the large-deviation statistics of the energy as we will see. These findings suggest an interesting duality between fluctuations in small systems and their phase behavior in the thermodynamic limit.

The paper is organized as follows. In Sec.~\ref{sec:zipper} we introduce the molecular zipper and show that it exhibits a thermal phase transition. In Sec.~\ref{sec:Lee-Yang} we demonstrate how the Lee-Yang zeros in the complex plane of the inverse temperature can be extracted from the energy cumulants for finite-size zippers. In Sec.~\ref{sec:LDF} we  discuss the connection between the Lee-Yang zeros and the large-deviation statistics. In Sec.~\ref{sec:broken_zipper} we introduce the broken zipper which exhibits a crossover rather than a sharp phase transition. Still, the Lee-Yang zeros can be extracted from the energy fluctuations. Moreover, in Sec.~\ref{sec:LDFandLY} we show that the large-deviation statistics takes the shape of an ellipse whose tilt and width are determined by the Lee-Yang zeros. An associated symmetry of the fluctuations is discussed in Sec.~\ref{sec:FR}. Finally, in Sec.~\ref{sec:conclusions} we present our conclusions and provide an outlook on future work. Technical details are deferred to the appendices.

\section{The Zipper Model}
\label{sec:zipper}

\begin{figure*}
	\centering
	\includegraphics[width=0.99\textwidth]{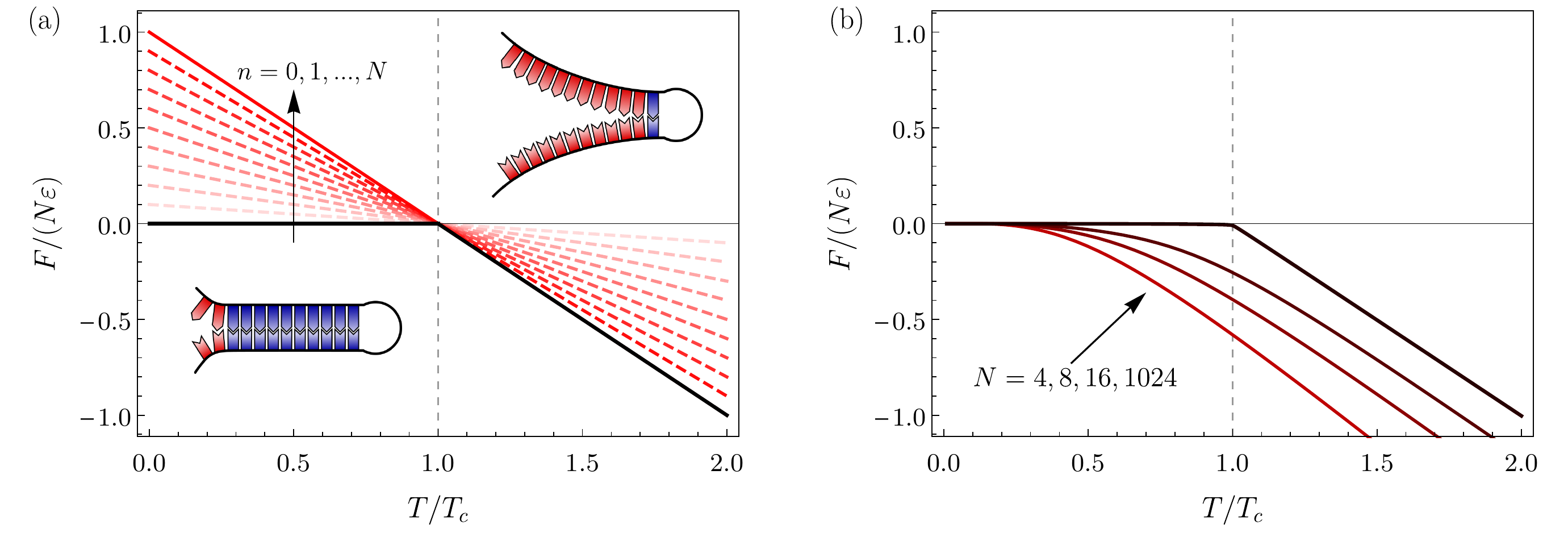}
	\caption{Free energy of the molecular zipper. (a) Free energy per link as a function of the temperature $T$ for different numbers of open links $n=0,1,\ldots,N$. For $T<T_c$, the closed zipper has the lowest free energy. For $T>T_c$, the open zipper has the lowest free energy. The equilibrium free energy is shown with a black line. (b) Equilibrium free energy per link for different lengths of the zipper. In the thermodynamic limit, the equilibrium free energy per link becomes non-analytic at $T=T_c$.}
\label{fig2}
\end{figure*}

Figure~\ref{fig1}~(a) shows the molecular zipper. The system consists of a double-stranded macromolecule held together by $N$ links that can be either open or closed.\cite{Kittel1969} If the zipper is fully closed, the internal energy $U$ is zero. The entropy $S$ is also zero, since there is only one configuration for the closed zipper, namely all links being closed. The free energy at temperature $T$
\begin{equation}
F=U-TS
\end{equation}
is then zero for the closed zipper. If $n\leq N$ links are open, the energy is $n\varepsilon$, since opening a link requires the energy $\varepsilon>0$. It is only possible to open a link if the preceding one is also open. Hence, links can only open up one after another starting from the floating end. This unwinding mechanism effectively leads to infinite-range interactions between distant links. Each link can open in $g$ different ways such that $n$ links can be open in $g^n$ different configurations with the corresponding entropy $S=k_B \ln g^n$, where $k_B$ is Boltzmann's constant. The free energy then becomes
\begin{equation}
F=n(\varepsilon-k_B T\ln g),\,\,\, 0\leq n\leq N.
\label{eq:free_energy}
\end{equation}
The free energy per link can be expressed as
\begin{equation}
\frac{F}{N\varepsilon}=\frac{n}{N}(1-T/T_c).
\label{eq:free_energy_diff}
\end{equation}
where we have defined the temperature
\begin{equation}
T_c=\frac{\varepsilon}{k_B \ln g}.
\end{equation}
The free energy per link corresponding to different numbers of open links is shown in Fig.~\ref{fig2} (a).

In equilibrium, the system minimizes the free energy. Thus, at low temperatures, where the free energy is dominated by the internal energy, the lowest free energy is zero corresponding to the closed zipper. With increasing temperature, the free energy becomes dominated by the entropy such that the free energy is now maximally negative for the fully open zipper. The transition between the two phases occurs at the temperature $T_c$, where the free energies are equal. The equilibrium free energy is also shown in Fig.~\ref{fig2} (a). We clearly see that the equilibrium free energy is non-analytic at $T=T_c$, where the free energy difference between the two phases vanishes. We note that there is no phase transition for $g=1$, where the temperature $T_c$ diverges.

The discussion above is based on simple thermodynamic arguments. A more rigorous analysis requires that we calculate the canonical partition function. For the zipper model, the partition function can easily be expressed as a sum over all possible numbers of open links
\begin{equation}\label{eq:PartFunct}
Z=\sum_{n=0}^{N} g^n e^{-\beta n \varepsilon }
=\frac{1 - (g e^{-\beta\varepsilon})^{N+1}}
{1-g e^{-\beta \varepsilon}},
\end{equation}
having introduced the inverse temperature $\beta= 1/(k_B T)$. The free energy for finite $N$ can then be obtained as
\begin{equation}
F=-\beta^{-1}\ln Z=-\beta^{-1}\ln \left[\frac{1 - (g e^{-\beta\varepsilon})^{N+1}}
{1-g e^{-\beta \varepsilon}}\right].
\label{eq:FreeEnerg}
\end{equation}
In Fig.~\ref{fig2} (b), we show the equilibrium free energy as a function of the temperature. With increasing length, the free energy per link approaches the thermodynamic value
\begin{equation}
\frac{F}{N\varepsilon}=\left\{
              \begin{array}{cc}
                1 -T/T_c, & T>T_c \\
                0, & T\leq T_c
              \end{array}
            \right..
\label{eq:free_energy_thermo}
\end{equation}
The two cases correspond exactly to the free energies of the open and the closed zipper, above and below the critical temperature, respectively. Interestingly, the zipper has to be rather long $N \sim 10^3$ before it becomes apparent that the free energy will become nonanalytic at $T_c$.

\begin{figure*}
	\centering
	\includegraphics[width=0.99\textwidth]{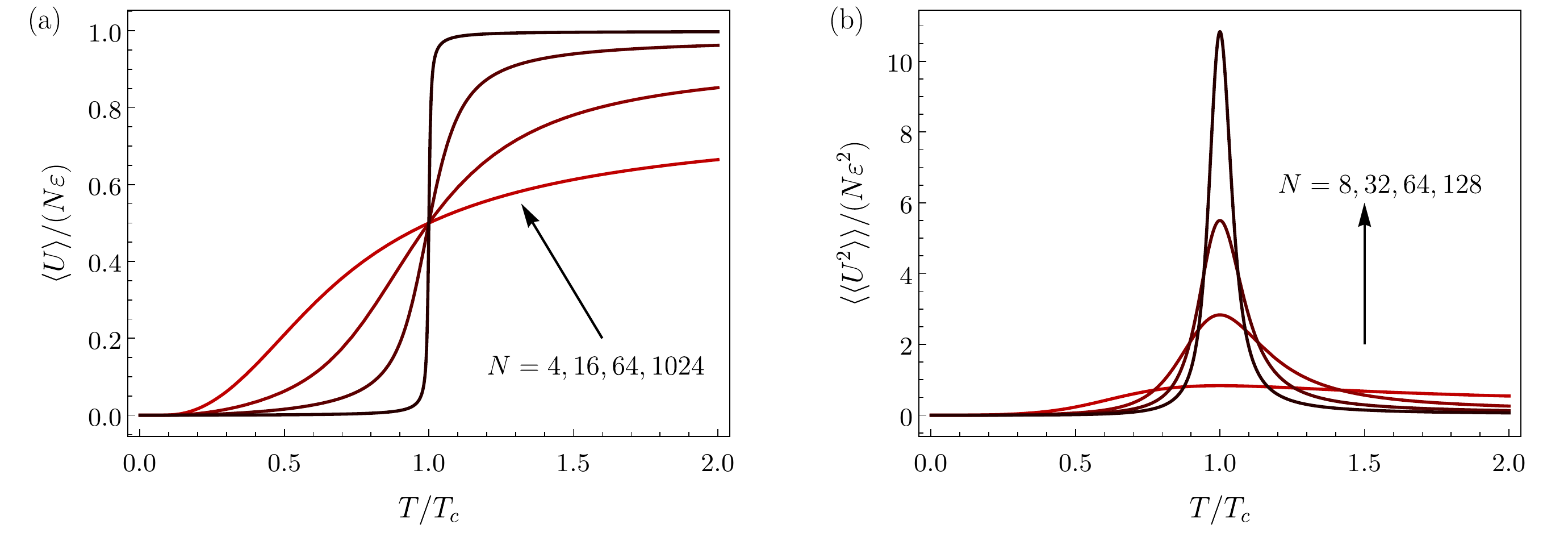}
	\caption{Average energy and fluctuations. (a) Average energy per link as a function of the temperature $T$. The various curves correspond to different lengths of the zipper. In the thermodynamic limit, the average energy per link develops a discontinuity at the  transition temperature $T_c$. (b) Variance of the energy for zippers of different lengths. The energy fluctuations diverge at the transition temperature $T_c$ in the thermodynamic limit.}
\label{fig3}
\end{figure*}

From the free energy, we find the average energy as $\langle U\rangle = \partial_\beta(\beta F)$. The average energy per link as a function of the temperature is shown in Fig.~\ref{fig3}~(a). Again, with increasing length, the average energy per link approaches the thermodynamic value
\begin{equation}
\frac{\langle U\rangle}{N\varepsilon}= \left\{
              \begin{array}{cc}
                1, & T>T_c \\
                1/2, & T=T_c \\
                0, & T<T_c
              \end{array}
            \right..
\end{equation}
This result shows that the system exhibits a first-order phase transition at $T=T_c$, where the average energy per link jumps abruptly from 0 to $\varepsilon$.

From the free energy, we also obtain the variance of the energy fluctuations as $\langle\!\langle U^2\rangle\!\rangle = -\partial^2_\beta(\beta F)$. In Fig.~\ref{fig3}~(b), we show the variance per link as a function of the temperature. Again, with increasing length, the variance approaches the thermodynamic value
\begin{equation}
\frac{\langle\!\langle U^2\rangle\!\rangle}{N\varepsilon^2}= \left\{
              \begin{array}{cc}
                0, & T\neq T_c \\
                \frac{N}{12}, & T=T_c\\
              \end{array}
            \right.,
\end{equation}
which diverges with the system size $N$ exactly at the critical temperature. Also for the average energy and the variance, the zipper has to be rather long before the phase transition becomes apparent.

\section{Lee-Yang zeros \& High cumulants}
\label{sec:Lee-Yang}

\begin{figure*}
	\centering
	\includegraphics[width=0.99\textwidth]{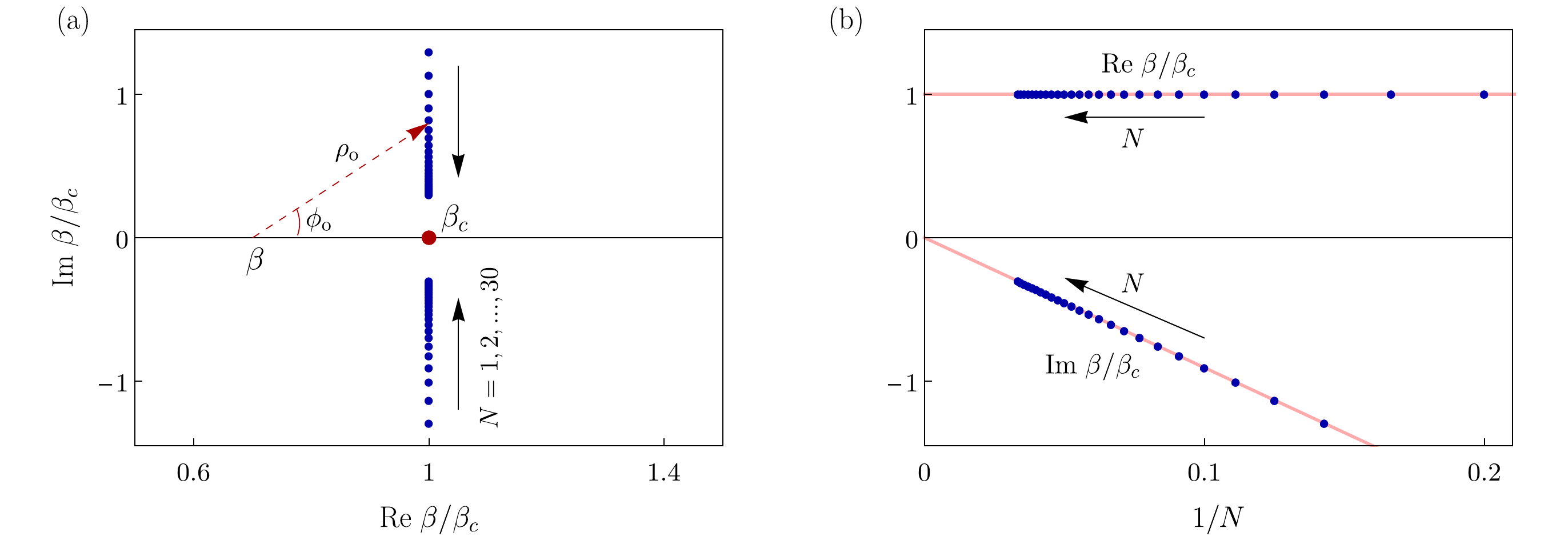}
	\caption{Lee-Yang zeros for the zipper model. (a) Exact results for the leading Lee-Yang zeros in the complex plane of the inverse temperature for zippers of increasing size, $N=1,2,...,30$. In the thermodynamic limit, the Lee-Yang zeros reach the inverse temperature on the real-axis for which a phase transition occurs. The polar coordinates $\rho_{\rm o}$ and $\phi_{\rm o}$ defined in Eq.~(\ref{eq:polar_coor}) are indicated. (b) Extrapolation of the real part and the imaginary part of the Lee-Yang zeros in the thermodynamic limit.}
\label{fig4}
\end{figure*}

To understand how the free energy can become nonanalytic in the thermodynamic limit, Lee and Yang suggested to investigate the zeros of the partition function in the complex plane of the control parameters.\cite{Yang1952a,Lee1952} In the case of the zipper model, we consider the zeros of the partition function for complex values of the inverse temperature $\beta$. (Zeros in the complex plane of the inverse temperature are sometimes referred to as Fisher zeros.\cite{fisher1965}) For real values of $\beta$, it is clear that the partition function in Eq.~(\ref{eq:PartFunct}) cannot vanish, since it is a sum of positive terms. On the other hand, treating $\beta$ as a complex variable, the partition sum vanishes at the points
\begin{equation}\label{eq:LYZStandard}
\beta_k = \beta_c + \frac{2 \pi k}{\varepsilon (N+1)}i,
\end{equation}
where $\beta_c=1/(k_BT_c)$ is the inverse transition temperature, and $k\in \{-N,\ldots,N\}\setminus\{0\}$ if we restrict ourselves to the stripe $-\pi\leq {{{\rm Im}}}\;\beta\varepsilon \leq \pi$ in the complex plane. In Fig.~\ref{fig4} (a), we show the motion of the Lee-Yang zeros that are closest to the real-axis with increasing system size. As expected from Eq.~(\ref{eq:LYZStandard}), the Lee-Yang zeros approach the inverse phase transition temperature $\beta_c$ in the thermodynamic limit. We note that the perpendicular approach to the real-axis signals that the system is exhibiting a first-order phase transition.\cite{grossmann1967,grossmann1969,Blythe2003} To illustrate how one may pinpoint the exact convergence point from finite-size systems, we show in Fig.~\ref{fig4} (b) the real part and the imaginary part of the leading Lee-Yang zeros as functions of the inverse system size. From such a figure, one can easily extrapolate the position of the Lee-Yang zeros in the large system-size limit.

Since the partition sum at finite size is an entire function, we can  write it in terms of the Lee-Yang zeros as\cite{arfken2012}
\begin{equation}
Z(\beta)=Z(0)e^{\beta c}\prod_k\left(1-\beta/\beta_k\right),
\end{equation}
where the constant $c=Z'(0)/Z(0)+\sum_k 1/\beta_k$ is independent of $\beta$. In general, it is challenging to determine the Lee-Yang zeros of a partition function and perform the product expansion. However, due to the simplicity of the zipper model, we can find all Lee-Yang zeros and explicitly construct the expansion as shown in App.~\ref{app:appB}.

The free energy can now be written as
\begin{equation}
F(\beta)=-\beta^{-1}\left[\ln Z(0) + \beta c +\sum_k\ln\left(1-\beta/\beta_k\right)\right].
\label{eq:free_energy_LY}
\end{equation}
This expression shows that the phase behavior of the system is fully encoded in the Lee-Yang zeros. We also see that the Lee-Yang zeros correspond to logarithmic singularities of the free energy. Moreover, as $N$ increases, the number of zeros grows and they approach the point $\beta=\beta_c$ on the real axis, where $F(\beta)$ becomes non-analytic. We note that the
partition function in Eq.~(\ref{eq:PartFunct}) does not vanish at $\beta_c$ in the limit $N\rightarrow\infty$. Rather, it becomes a discontinuous function along the vertical line in the complex plane that crosses the real-axis at $\beta=\beta_c$.

Lee-Yang zeros are useful to understand phase transitions. However, until recently, it has been unclear how to determine Lee-Yang zeros in experiment. As we will now discuss, the Lee-Yang zeros can be obtained from the fluctuations of thermodynamic observables of finite-size systems.\cite{flindt2013,Brandner2017,hickey2013,hickey2014} To this end, we first note that the partition function delivers the moments of the energy upon differentiation with respect to the inverse temperature as
\begin{equation}
\langle U^n\rangle =[(-\partial_{\beta})^nZ]/Z.
\label{eq:moments}
\end{equation}
Similarly, the cumulants can be obtained as
\begin{equation}
\langle\!\langle U^n\rangle\!\rangle=(-\partial_{\beta})^n \ln Z=(-\partial_{\beta})^n (-\beta F).
\label{eq:cumulants}
\end{equation}
Now, using Eq.~(\ref{eq:free_energy_LY}), we can express the cumulants as
\begin{equation}
\langle\!\langle U^n\rangle\!\rangle= (-1)^{n-1} \sum_{k} \frac{(n-1)!}{(\beta_k-\beta)^n},\,\, n>1.
\end{equation}
The fact that the Lee-Yang zeros come in complex-conjugate pairs ensures that the cumulants are real.

The result above shows that the energy fluctuations are fully described by the Lee-Yang zeros. Moreover, we see that the high cumulants essentially are governed by the pair of Lee-Yang zeros that are closest to the inverse temperature $\beta$ on the real-axis, since they dominate the sum for high orders, $n\gg1$. The contributions from sub-leading zeros are suppressed with the distance to $\beta$ and the order $n$. Now, writing the closest pair of Lee-Yang zeros, $\beta_{\rm o}$ and $\beta_{\rm o}^*$, in polar coordinates as, see Fig.~\ref{fig4}~(a),
\begin{equation}
\beta_{\rm o} = \rho_{\rm o} e^{i\phi_{\rm o}}+\beta,
\label{eq:polar_coor}
\end{equation}
we can approximate the sum as
\begin{equation}
\langle\!\langle U^n\rangle\!\rangle\simeq (-1)^{n-1}(n-1)!\frac{2 \cos (n \phi_{\rm o})}{\rho_{\rm o}^n},\,\, n\gg1.
\label{eq:high_cumu}
\end{equation}
The approximation becomes better with increasing order and it shows how the high cumulants are determined by the closest Lee-Yang zeros. We also see that the high cumulants should oscillate for example as a function of the order $n$ or the inverse temperature $\beta$. Such oscillations are generic features of high derivatives\cite{dingle1973,berry2005,flindt2010} and they have been observed in measurements of the high cumulants of the electron counting statistics in quantum dots.\cite{flindt2009,fricke2010,fricke2010b,kambly2011}

Equation (\ref{eq:high_cumu}) is important as it relates the Lee-Yang zeros to experimental observables. In particular, by measuring the high cumulants of the energy, one can extract the position of the leading pair of Lee-Yang zeros. To this end, we solve Eq.~(\ref{eq:high_cumu}) for $\rho_o$ and $\phi_o$ and find\cite{zamastil2005,flindt2010,kambly2011,flindt2013}
\begin{equation}
\label{eq:cumMethod}
\begin{bmatrix}
1& \frac{-\kappa_{n}^{(+)}}{n}\\
1& \frac{-\kappa_{n+1}^{(+)}}{n+1}
\end{bmatrix}
\begin{bmatrix}
-2\rho_{\rm o} \cos \phi_{\rm o}\\
\rho_{\rm o}^2
\end{bmatrix}=\begin{bmatrix}
(n-1) \kappa_{n}^{(-)}\\
n \ \kappa_{n+1}^{(-)}\\
\end{bmatrix}
\end{equation}
in terms of the ratios of cumulants
\begin{equation}
\kappa_{n}^{(\pm)} \equiv \langle\!\langle U^{n\pm 1}\rangle\!\rangle / \langle\!\langle U^{n}\rangle\!\rangle.
\end{equation}
Thus, from four consecutive cumulants, one may extract the leading pair of Lee-Yang zeros by solving Eq.~(\ref{eq:cumMethod}).

\begin{figure*}
	\centering
	\includegraphics[width=0.99\textwidth]{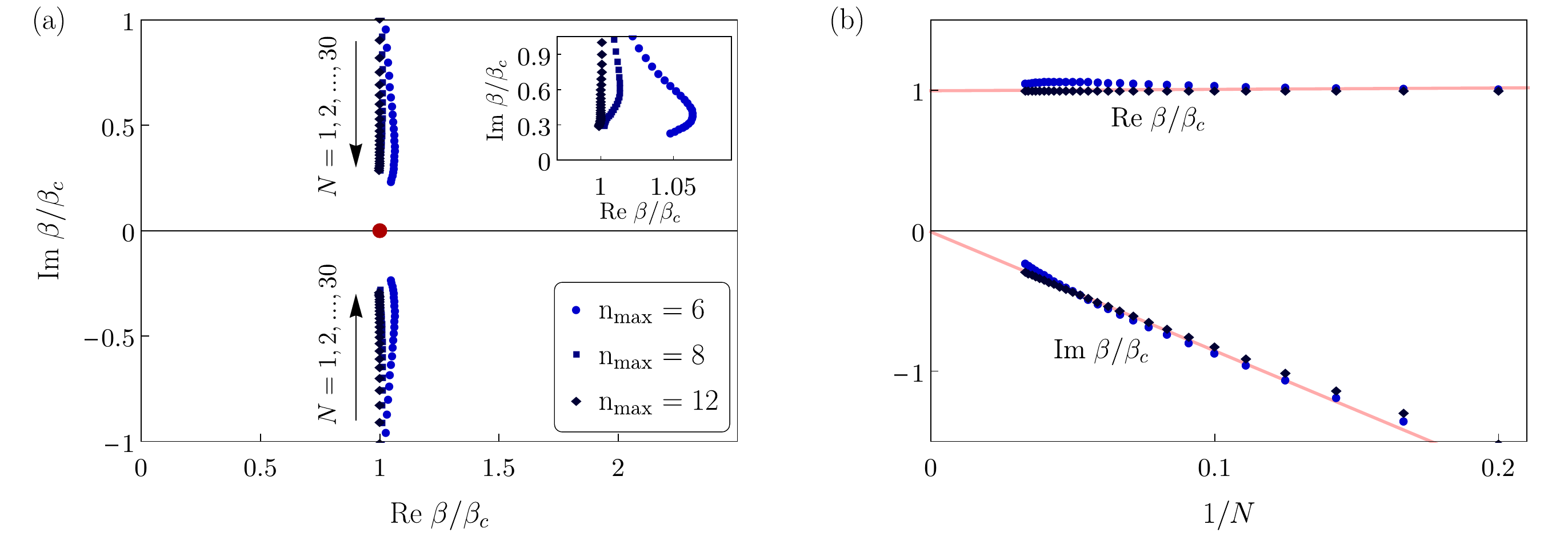}
	\caption{Extraction of Lee-Yang zeros from the energy cumulants. (a) The Lee-Yang zeros are extracted from four consecutive cumulants up to order $n_\mathrm{max}$ for zippers of length $N=1,2,\ldots,30$. The inverse temperature is $\beta=0.75 \beta_c$. The inset illustrates how the accuracy of the method is improved by increasing the cumulant order. (b) Extrapolation of the real and imaginary part of the Lee-Yang zeros in the thermodynamic limit. These results should be compared with the exact ones in Fig.~\ref{fig4}.}
\label{fig5}
\end{figure*}

To illustrate the method, we extract the leading Lee-Yang zeros for the zipper model using the cumulants of the energy. In Fig.~\ref{fig5} (a), we show the motion of the leading Lee-Yang zeros in the complex plane of the inverse temperature. With increasing system size, the Lee-Yang zeros move towards the real-axis. In Fig.~\ref{fig5} (b), we again show the real part and the imaginary part of the Lee-Yang zeros as functions of the inverse system size, allowing us to extract the exact convergence point. We see that the accuracy of the method improves with the order of the energy cumulants. We note that the method is not restricted to energy cumulants and Lee-Yang zeros in the complex plane of the inverse temperature. Other sets of conjugate variables would also work.

\section{Large-deviation statistics}
\label{sec:LDF}

Having illustrated our method for determining the Lee-Yang zeros from experimental observables, we now go on to discuss the implications for the large-deviation statistics of the energy.\cite{touchette2009} To this end, we evaluate the probability $P(u)$ for finding the zipper with the energy $u=n\varepsilon/N$ per link. We can express this probability as
\begin{equation}
P(u)=\frac{g^n e^{-\beta Nu}}{Z}
\end{equation}
with the partition function $Z$ given by Eq.~(\ref{eq:PartFunct}). We focus on the tails of the distribution by considering the logarithm of the probability distribution
\begin{equation}
\frac{\ln P(u)}{N}=u\left(\beta_c-\beta\right) +\frac{\beta F}{N}.
\end{equation}
For long zippers, $N\gg1$, we can use Eq.~(\ref{eq:free_energy_thermo}) for the free energy, and we then find the simple expression
\begin{equation}
\label{eq:ldf_real}
\frac{\ln P(u)}{N}=
\begin{cases}
(\beta_{c}-\beta) u, & \beta \geq \beta_{c} \\
(\beta_{c}-\beta)(u-\varepsilon), & \beta < \beta_{c}
\end{cases},
\end{equation}
which vanishes exactly at the phase transition $\beta=\beta_c$. In Fig.~\ref{fig1} (c), we show the large-deviation function for temperatures below, equal to, or above the phase transition temperature. The large-deviation function describes the exponential decay of the probability to observe energy fluctuations away from the average value. Equation~(\ref{eq:ldf_real}) shows that the large-deviation function is determined by the inverse transition temperature $\beta_c$.  As we will see below, this behavior is of a more general nature, namely, the convergence points of the Lee-Yang zeros determine the large-deviation statistics of thermodynamic observables.

Before closing this section, we briefly discuss the cumulants of the energy in the thermodynamic limit. Away from the phase transition, the free energy and all cumulants of the energy grow linearly with the system size~$N$. On the other hand, at the phase transition we find
\begin{equation}
\langle\!\langle U^n\rangle\!\rangle =\varepsilon^n\zeta(1-n)(1-N^n),
\end{equation}
where $\zeta(x)$ is the Riemann zeta function which for the first few cumulants takes the values $\zeta(0)=-1/2$, $\zeta(-1)=-1/12$, $\zeta(-2)=0$, and $\zeta(-3)=-1/120$. This particular behavior of the cumulants is closely related to the strongly non-gaussian energy fluctuations at the phase transition. Specifically, if we introduce the variable
\begin{equation}
\vartheta=\frac{U-\langle U\rangle}{\sqrt{\langle\!\langle U^2\rangle\!\rangle}}
\end{equation}
which measures deviations from the average energy in units given by the variance,\cite{karzig2010} we find $\langle \vartheta \rangle = 0$, $\langle\!\langle\vartheta^2\rangle\!\rangle=1$, and $\langle\!\langle\vartheta^n\rangle\!\rangle\simeq0$ for $n>2$, away from the phase transition. This shows that $\vartheta$ essentially is governed by a Gaussian distribution with only small deviations in the tails. By contrast, at the phase transition, we find
\begin{equation}
\langle\!\langle\vartheta^n\rangle\!\rangle=-\zeta(1-n)(2\sqrt{3})^{n},\,\, n\geq 2
\end{equation}
in the thermodynamic limit. Thus, the scaled high-order cumulants are non-zero, reflecting the highly non-Gaussian fluctuation of $\vartheta$ at the phase transition.

\section{Broken Zipper}
\label{sec:broken_zipper}

\begin{figure*}
\centering
\includegraphics[width=0.99\textwidth]{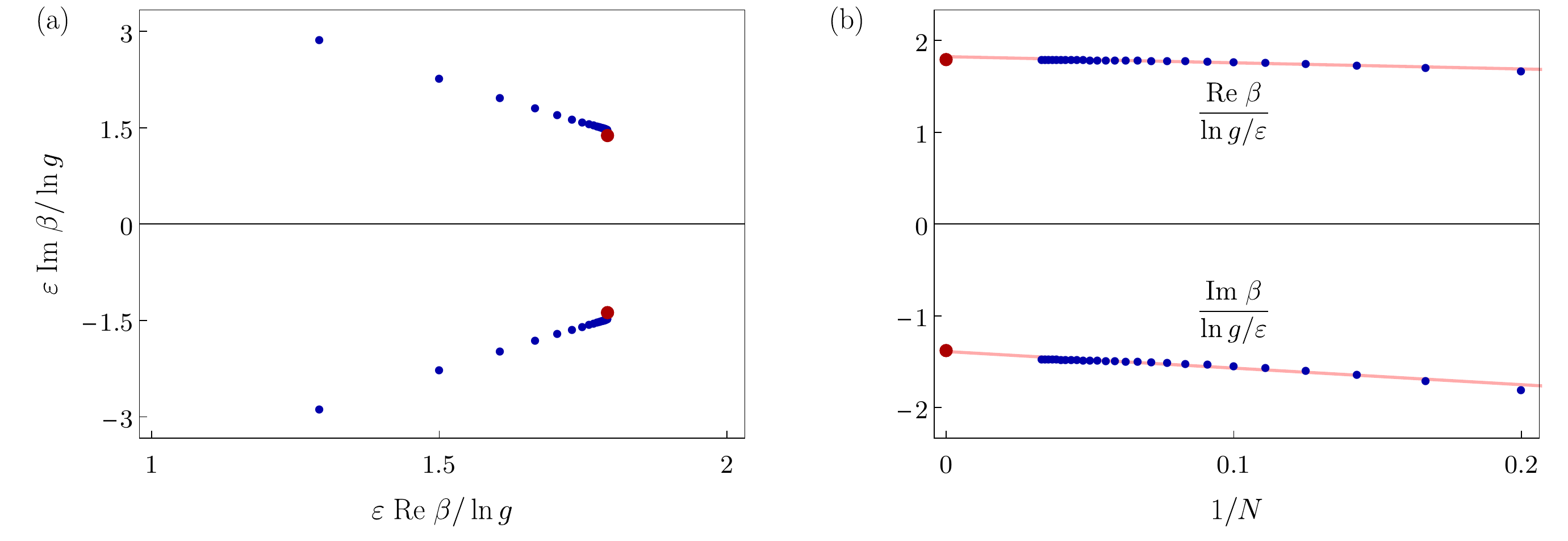}
\caption{Lee-Yang zeros for the broken zipper. (a) Leading Lee-Yang zeros in the complex plane of the inverse temperature. The zeros are extracted from the energy cumulants $\langle\!\langle U^n\rangle\!\rangle$ of order $n=9,10,11,12$ for zippers of length $N=1,2,\ldots,30$ with $J=\varepsilon$ and $\beta=1.75 (\ln g)/\varepsilon$. With increasing size, the Lee-Yang zeros move towards the convergence points marked with red circles. (b)~Extrapolation of the convergence points $\beta_c$ and $\beta_c^*$ in the thermodynamic limit. The Lee-Yang zeros remain complex.}
\label{fig:Lee-Yang_broken}
\end{figure*}

We are now ready to illustrate our method with a more evolved model. To this end, we consider finite values of the energy cost $J$ associated with opening a link inside a closed part of the zipper. In this case, the total energy of the system can be expressed as\cite{cuesta2004}
\begin{equation}
U=\varepsilon \sum_{i=1}^N \sigma_{i} +J\sum_{i=2}^{N} \sigma_{i} (1-\sigma_{i-1}),
\end{equation}
where $\sigma_{i}=1$ for open links and $\sigma_{i}=0$ for closed links. The corresponding partition function reads
\begin{equation}
Z=\sum_{\{\sigma_{i}\}} g(\{\sigma_{i}\}) \ e^{-\beta U(\{\sigma_{i}\})},
\end{equation}
where the degeneracy factor associated with a particular sequence $\{\sigma_{i}\}$ is
\begin{equation}
 g(\{\sigma_{i}\})=g^{\sum_{i}^{N} \sigma_{i}},
\end{equation}
recalling that each link can be open in $g>1$ ways. The partition function can now be written as
\begin{equation}
Z= \sum_{\{\sigma_{i}\}} g^{\sigma_{1}} e^{-\beta\varepsilon\sigma_{1}} \prod_{i=2}^{N} T(\sigma_{i},\sigma_{i-1}),
\end{equation}
where
\begin{equation}
T(\sigma_{i},\sigma_{i-1})=g^{\sigma_{i}} e^{-\beta (J + \varepsilon){\sigma_{i}}}e^{\beta J \sigma_{i}\sigma_{i-1}}
\end{equation}
are elements of the transfer matrix
\begin{equation}
\label{eq:transferMatrix}
\mathbf{T}=\begin{pmatrix}
g e^{-\beta\varepsilon}&g e^{-\beta( J +\varepsilon)}\\
1&1
\end{pmatrix}.
\end{equation}
Imposing the appropriate boundary conditions, the partition sum can be expressed in the compact form
\begin{equation}
\label{eq:Ztrans}
Z=\langle -|\mathbf{T}^N| 1\rangle
\end{equation}
with $\langle -|=(1,1)$ and $|1\rangle=(1,0)^T$.  This expression allows us to calculate the partition function and the free energy for zippers of finite size together with the cumulants of the energy.
The partition function in Eq.~(\ref{eq:PartFunct}) is readily recovered in the limit $J \rightarrow \infty$. From Eq.~(\ref{eq:Ztrans}), we also see that the partition sum must take the form
\begin{equation}
Z=c_1\lambda_1^N+c_2\lambda_2^N,
\end{equation}
where $\lambda_1$ and $\lambda_2$ are the eigenvalues of the transfer matrix and the prefactors $c_1$ and $c_2$ are independent of the system size. In the thermodynamic limit, the free energy
\begin{equation}
\frac{F}{N}=-k_BT\ln \lambda_\mathrm{max}
\end{equation}
is thus given by the eigenvalue of the transfer matrix with the largest absolute value. Correspondingly, we see that non-analyticities in the free energy, associated with phase transitions, are given by eigenvalue crossings of the transfer matrix, i.e.~values of $\beta$ for which $\lambda_1=\lambda_2$. We also see that the Lee-Yang zeros with increasing system size approach these values of $\beta$. Specifically, by solving for the zeros of the partition function, we find
\begin{equation}
\ln(\lambda_1)=\ln(\lambda_2)+\frac{\ln(c_2/c_1)+i\pi(2k+1)}{N}
\label{eq:eq4zeros}
\end{equation}
for integer $k$. Thus, the Lee-Yang zeros will move towards the values of $\beta$ for which $\lambda_1=\lambda_2$, since the second term on the righthand side vanishes for $N\rightarrow\infty$. This conclusion holds for transfer matrices of any size, where the phase transition is determined by a crossing of the two largest eigenvalues.

\begin{figure*}
	\centering
	\includegraphics[width=1\textwidth]{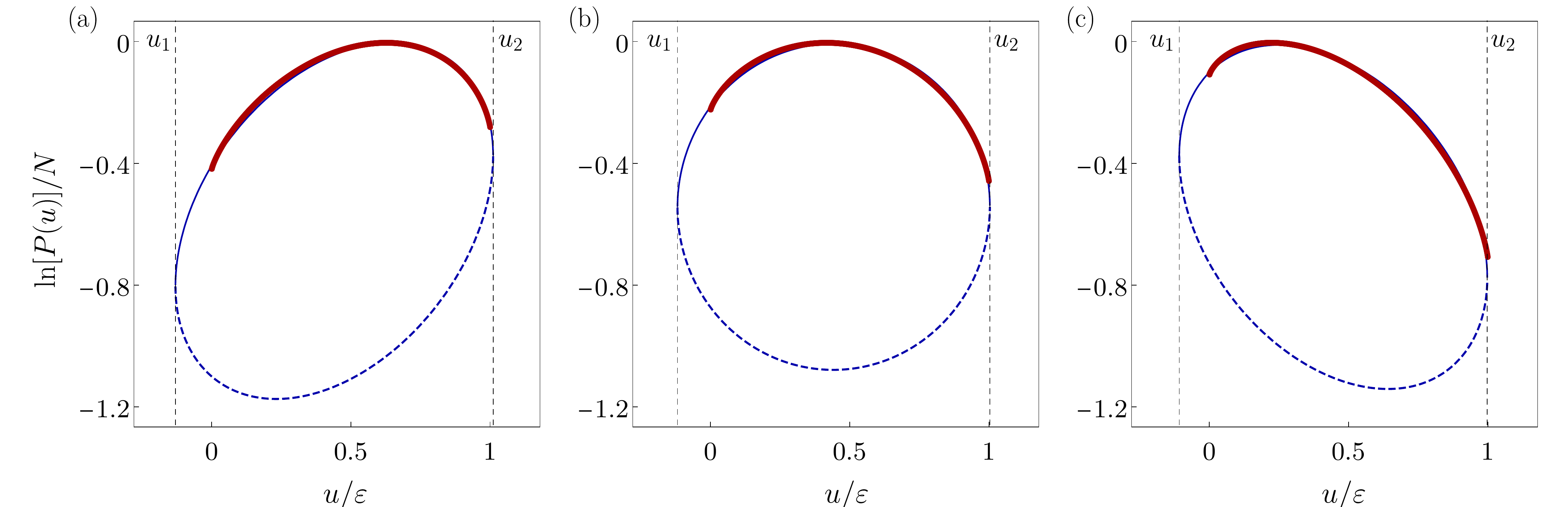}
	\caption{\label{fig:LDFBroken}  Large-deviation statistics for the broken zipper. The inverse temperature is (a) $\beta=0.7\, \mathrm{Re[\beta_c]}$, (b) $\beta=\mathrm{Re[\beta_c]}$, and (c) $\beta=1.3\,\mathrm{Re[\beta_c]}$, where $\mathrm{Re[\beta_c]}$ is the real part of the convergence points extracted in Fig.~\ref{fig:Lee-Yang_broken} with $J=\varepsilon$. The red lines are exact results based on Eq.~(\ref{eq:LDF_exact}) with $N=10^3$, while the blue lines are the top of the tilted ellipse described by Eq.~(\ref{eq:ldfgen}). The fitting parameters $u_1$, $u_2$, and $\mathcal{N}$  depend weakly on the temperature. The lowers parts of the ellipses are indicated with dashed lines.}
\end{figure*}

The two eigenvalues of the transfer matrix are
\begin{equation}
\lambda_{1,2}=\frac{1+ g e^{-\beta\varepsilon}}{2}  \pm  \sqrt{\left(\frac{1-g e^{-\beta\varepsilon}}{2}\right)^2+ g e^{-\beta(\varepsilon+J)}}.
\end{equation}
An eigenvalue crossing occurs when the argument of the square-root is zero. For real values of $\beta$, this can only happen with an infinitely large $J$. In that case, the system exhibits a phase transition for $\beta=(\ln g)/\varepsilon$ as we saw earlier on. For finite values of $J$, eigenvalue crossings can only occur for complex values of $\beta$ and the system does not exhibit a sharp phase transition at any temperature.

We find the Lee-Yang zeros from the high cumulants of the energy using the extraction scheme in Eq.~(\ref{eq:cumMethod}). An example of this procedure is shown in Fig.~\ref{fig:Lee-Yang_broken}~(a). The dominating Lee-Yang zeros are extracted for short zippers of a finite length. To find the convergence points of the Lee-Yang zeros, we need to extrapolate their position in the thermodynamic limit. To this end, we note that $1/N$ appears as a natural expansion parameter, e.g. in Eq.~(\ref{eq:eq4zeros}). We therefore expand the Lee-Yang zeros as
\begin{equation}
\begin{split}
\mathrm{Re}[\beta_\mathrm{o}]&=\mathrm{Re}[\beta_c]+\alpha_1^\mathrm{Re}/N+\alpha_2^\mathrm{Re}/N^2+\ldots\\
\mathrm{Im}[\beta_\mathrm{o}]&=\mathrm{Im}[\beta_c]+\alpha_1^\mathrm{Im}/N+\alpha_2^\mathrm{Im}/N^2+\ldots
\end{split}
\end{equation}
with unknown expansion parameters. With an infinitely large $J$, we can indeed write the dominating Lee-Yang zeros from Eq.~(\ref{eq:LYZStandard}) as $\mathrm{Re}[\beta_\mathrm{o}]=\mathrm{Re}[\beta_c]$ and $\mathrm{Im}[\beta_\mathrm{o}]=\pm(2\pi/\varepsilon)(1/N-1/N^2+1/N^3+\ldots)$. For finite values of $J$ we assume that such a system-size dependence still holds. We can then use linear interpolation in $1/N$ with increasing system size. In Fig.~\ref{fig:Lee-Yang_broken}~(b) we show the extraction of the convergence points in the thermodynamic limit. We see that the Lee-Yang zeros in this case remain complex corresponding to a cross-over rather than a sharp phase transition. However, as we will now see, the Lee-Yang zeros still carry important information about the large-deviation statistics and its symmetry properties.

\section{Large deviations \& Lee-Yang zeros}
\label{sec:LDFandLY}

We are now ready to connect the Lee-Yang zeros with the large-deviation statistics of the energy fluctuations. To this end, we first write the partition sum as
\begin{equation}
Z=\sum_{n=0}^{N}\sum_{m=0}^{N-n} g^n\binom{n}{m}\binom{N-n}{m} e^{-\beta (n\varepsilon  + m J )}.
\end{equation}
Here, the energy and degeneracy factor associated with having $n$ open links are $n\varepsilon$ and $g^n$, respectively, and $m J$ is the energy associated with having $m$ open segments inside the zipper. The corresponding degeneracy factor is given by the product of binomial coefficients which follows from simple combinatorial arguments.

The probability density of finding the zipper with energy $u=U/N$ per link can be written as
\begin{equation}
P(u)=N\sum_{n=0}^{N}\sum_{m=0}^{N-n}\frac{g^n \binom{n}{m}\binom{N-n}{m}e^{-\beta Nu}}{Z} \delta (n\varepsilon  + m J-Nu).
\label{eq:LDF_exact}
\end{equation}
This expressions forms the basis of our exact calculations of the large-deviation statistics of the energy fluctuations. However, to connect the large-deviation statistics to the Lee-Yang zeros, we proceed with an approximation for large systems. Specifically, for $N\gg1$, we can approximate the probability density as
\begin{equation}
P(u)\simeq \frac{N}{2 \pi} \int_{-\infty}^{+\infty} d\gamma \frac{Z(\beta+i \gamma)}{Z(\beta)}e^{i \gamma N u}
\end{equation}
as shown in App.~\ref{app:appA}. For large systems, the partition sum is determined by the largest eigenvalue of the transfer matrix such that
\begin{equation}
P(u)\simeq \frac{N}{2 \pi} \int_{-\infty}^{+\infty} d\gamma \left[\frac{\lambda_{\max}(\beta+i \gamma)}{\lambda_{\max}(\beta)}\right]^Ne^{i \gamma N u}.
\end{equation}
Now, via a change of variable, we obtain the expression
\begin{equation}
P(u)\simeq \frac{N }{2 \pi i } \int_{\beta-i\infty}^{\beta+i\infty} d\gamma e^{N \left[(\gamma-\beta) u+\ln \left(\frac{\lambda_{\max}(\gamma)}{\lambda_{\max}(\beta)}\right)\right]},
\end{equation}
which is suitable for a saddle-point approximation in the large-$N$ limit. To find the saddle-points, we need to examine the exponent of the integrand and we thus define
\begin{equation}
\Theta(\gamma)=(\gamma-\beta) u+\ln \left(\frac{\lambda_{\max}(\gamma)}{\lambda_{\max}(\beta)}\right).
\end{equation}
The saddle-points are given by values of $\gamma$ for which the first derivative of $\Theta(\gamma)$ vanishes, i.e.~$\Theta'(\gamma_0)=0$. To find the saddle-points, we note that the integral should be dominated by the square-root branch points of $\lambda_{\max}(\gamma)$ closest to the real axis. Hence, we make the ansatz
\begin{equation}
\Theta(\gamma)\simeq (\gamma-\beta) u + c_1 + c_2 \gamma \pm c_3 \sqrt{(\beta_{c}-\gamma)(\beta_{c}^{*}-\gamma)}
\label{eq:ansatz}
\end{equation}
with plus for $\mathrm{Re}[\gamma] < \mathrm{Re}[\beta_{c}]$ and minus for $\mathrm{Re}[\gamma] > \mathrm{Re}[\beta_{c}]$. Here, $\beta_{c}$ and $\beta_{c}^{*}$ are the convergence points of the Lee-Yang zeros in the thermodynamic limit, and $c_1$, $c_2$, and $c_3$ are unknown constants. This ansatz captures the analytic structure of $\Theta(\gamma)$ around $\beta_{c}$ and $\beta_{c}^{*}$.

The first derivative of $\Theta(\gamma)$ now becomes
\begin{equation}
\Theta'(\gamma)= u + c_2 \pm c_3 \frac{\gamma-\mathrm{Re}[\beta_{c}]}{\sqrt{(\gamma-\mathrm{Re}[\beta_{c}])^2+\mathrm{Im}[\beta_{c}]^2}}.
\end{equation}
For the saddle-points, we then find
\begin{equation}
\gamma_0(u)=\mathrm{Re}[\beta_{c}] \pm \mathrm{Im}[\beta_{c}]\frac{ |u+c_2|}{\sqrt{c_3^2 - (u+c_2)^2}},
\end{equation}
where the sign should be chosen consistently with Eq.~(\ref{eq:ansatz}). Inserting this result back into the ansatz and defining $\mathcal{M}=c_1 + c_2  \mathrm{Re}[\beta_{c}]$ and $u_{1,2}=-c_2\pm c_3$, we find for the large-deviation function $\ln[ P(u)]/N\simeq \Theta(\gamma_0)$
\begin{equation}
\label{eq:ellipse}
\frac{\ln P(u)}{N}\simeq \mathcal{M} +(\mathrm{Re}[\beta_{c}]- \beta) u + \mathrm{Im}[\beta_{c}] \sqrt{(u_2-u)(u-u_1)}.
\end{equation}
This result generalizes Eq.~(\ref{eq:ldf_real}) to the case, where the convergence point $\beta_{c}$ is complex. Equation~(\ref{eq:ldf_real}) is recovered for $\mathrm{Im}[\beta_{c}]=0$ with an appropriate choice of~$\mathcal{M}$.

The large-deviation function can also be written as
\begin{equation}
\label{eq:ldfgen}
\frac{\ln P(u)}{N}\simeq\mathcal{N}-\beta u -\frac{1}{2}\!
\left(\!\sqrt{\beta_+(u\!-\!u_2)}
     -\sqrt{\beta_{-}(u\!-\!u_1)}
     \right)^2\!\!\!\!,
\end{equation}
for $u_1\leq u\leq u_2$, where $\mathcal{N}$ is a normalization constant and we have defined $\beta_{\pm}=|\beta_{c}|\!\pm\!\mathrm{Re}[\beta_{c}]$. This result generalizes the expression from Ref.~\onlinecite{Brandner2017} by including a non-zero control parameter, here the inverse temperature~$\beta$. Interestingly, the shape of the large-deviation function is given by the upper part of a tilted ellipse.\cite{jordan2004,lambert2015,singh2016,Brandner2017} The tilt is determined by the difference between the real part of the convergence point $\mathrm{Re}[\beta_{c}]$ and the inverse temperature $\beta$. The width is mainly controlled by the imaginary part $\mathrm{Im}[\beta_{c}]$. Thus, from the convergence points we can predict the shape and the tilt of the ellipse. The vertical off-set $\mathcal{N}$ together with $u_1$ and $u_2$ must be determined by fitting the ellipse to the exact large-deviation function.

In Fig.~\ref{fig:LDFBroken} we show the large-deviation statistics for three different temperatures. Together with the exact results, we show the ellipse given by Eq.~(\ref{eq:ellipse}) with the convergence points extracted in Fig.~(\ref{fig:Lee-Yang_broken}). We see how the Lee-Yang zeros fully capture the large-deviation statistics with the tilt of the ellipse controlled by the temperature. The fitting parameters $u_1$, $u_2$, and $\mathcal{N}$ are only weakly dependent on the temperature. This dependence comes about  due to the ansatz in Eq.~(\ref{eq:ansatz}) which only retains the linear dependence on the inverse temperature. The weaker temperature dependence is encoded in the parameters $c_1$, $c_2$, and $c_3$. It is remarkable how the large-deviation statistics can be related to the Lee-Yang zeros. Moreover, the assumption that the saddle-points are given by the square-root branch-points of the largest eigenvalue of the transfer matrix may be valid in many cases. We thus expect that the relation between Lee-Yang zeros and the large-deviation statistics may hold for many thermodynamic systems with co-existing phases.

\begin{figure}
	\centering
	\includegraphics[width=0.9\columnwidth]{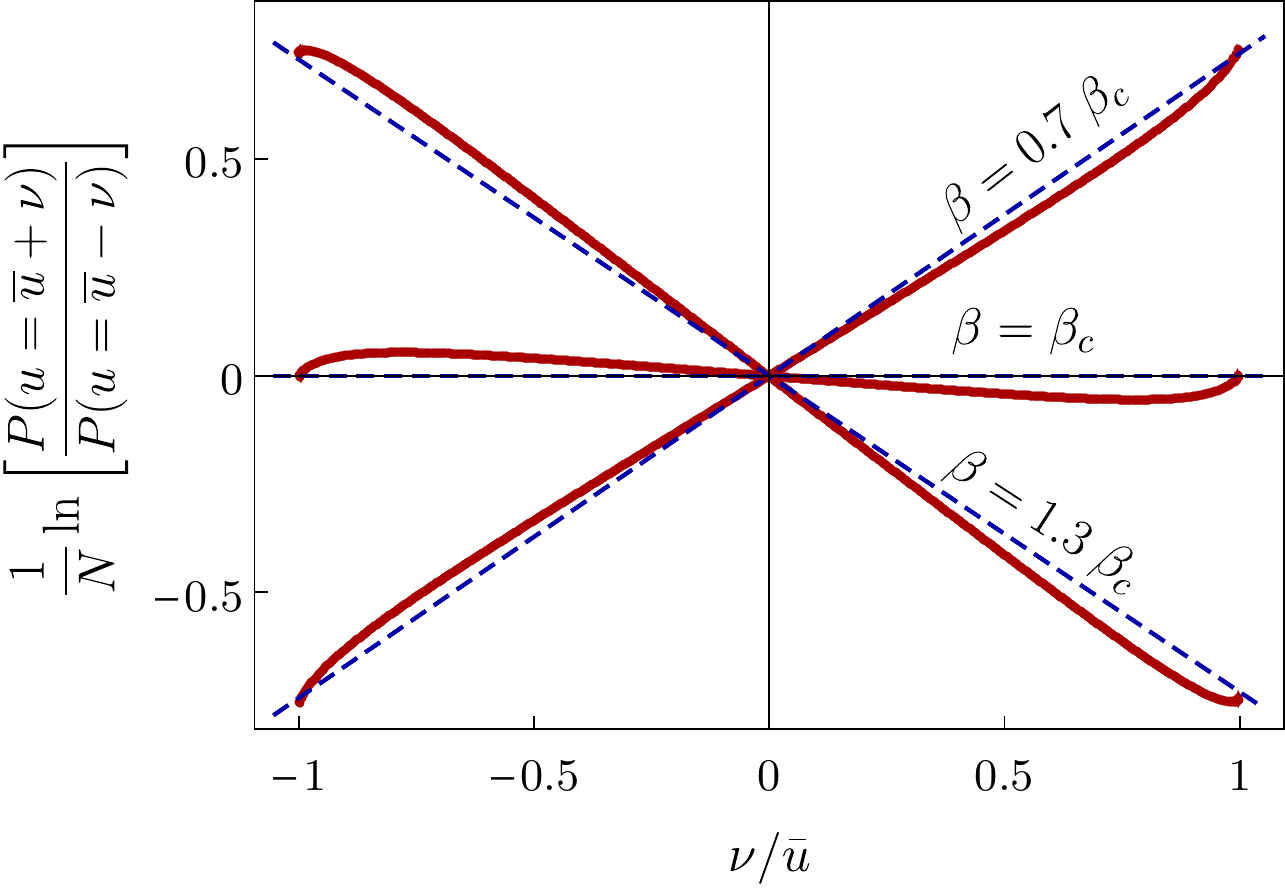}
	\caption{\label{fig:fluctuation} Fluctuation relation. The full lines show the lefthand side of Eq.~(\ref{eq:FT}) using the exact results from Fig.~\ref{fig:LDFBroken}. The dashed lines correspond to the righthand side of Eq.~(\ref{eq:FT}).}
\end{figure}

\section{Fluctuation relation}
\label{sec:FR}

Finally, we discuss the symmetry properties of the large-deviation statistics.\cite{esposito2009} In a spirit similar to that of Ref.~\onlinecite{singh2016}, it follows easily from Eq.~(\ref{eq:ellipse}) that the fluctuations for large systems obey the symmetry relation
\begin{equation}
\frac{1}{N}\ln\left[\frac{P(u=\bar{u}+\nu)}{P(u=\bar{u}-\nu)}\right]=2(\mathrm{Re}[\beta_{c}]- \beta)\nu,
\label{eq:FT}
\end{equation}
where we have introduced the average $\bar{u}=(u_1+u_2)/2$. This relation resembles the Gallavotti-Cohen fluctuation theorem from non-equilibrium statistical mechanics.\cite{gallavotti1995,gallavotti1995b} However, rather than the entropy production, this fluctuation symmetry concerns the departure $\nu$ from the average value $\bar{u}$. The slope on the righthand side is fully determined by the real-part of the convergence point minus the inverse temperature. Equation (\ref{eq:FT}) has a clear physical interpretation. For $\beta = \mathrm{Re}[\beta_{c}]$, fluctuations away from $\bar{u}$ are equally likely in both directions. Hence, the system has no tendency to settle either to the closed or the open phase. Instead, the microstate of the system is determined by large fluctuations which increase as the Lee-Yang zeros approach the real-axis. For $\beta > \mathrm{Re}[\beta_{c}]$, the probability of fluctuations that involve the opening of links is exponentially suppressed compared to the probability of fluctuations, where links are closed. Thus, the system is forced into the closed state and any departure from it is exponentially unlikely. The analogous argument holds for $\beta < \mathrm{Re}[\beta_{c}]$. In Fig.~\ref{fig:fluctuation}, we check the expected fluctuation symmetry using the results from Fig.~(\ref{fig:LDFBroken}) and find good agreement with Eq.~(\ref{eq:FT}). This finding is suggestive of an intimate connection between Lee-Yang zeros and fluctuation relations, both in equilibrium and non-equilibrium situations.

\section{Conclusions}
\label{sec:conclusions}

We have investigated a thermal phase transition in a molecular zipper using a recently established method to extract the dominant Lee-Yang zeros from fluctuations of thermodynamic observables at finite size. In the present case, we extracted the Lee-Yang zeros in the complex plane of the inverse temperature from the fluctuations of the energy. Other conjugate variables in classical systems would also work. For instance in a magnetic system, the Lee-Yang zeros in the complex plane of the magnetic field can be extracted from the fluctuations of the magnetization. We have shown that we can predict the temperature at which a phase transition occurs in the thermodynamic limit from the energy fluctuations in small zippers, also at temperatures well above (or below) the transition point. Even if the system does not undergo a sharp transition, the Lee-Yang zeros carry important information about the large-deviation statistics and its symmetry properties. Our work suggests an interesting duality between fluctuations in small systems and their phase behavior in the thermodynamic limit. These predictions may be tested in future experiments on scalable many-body systems. Moreover, the general framework presented here is not restricted to equilibrium systems. It can also be applied to non-conventional phase transitions such as dynamical phase transitions in quantum systems after a quench or dynamic order-disorder transitions in glasses. Adapting these ideas to quantum phase transitions constitutes another interesting line of research.

\begin{acknowledgments}
We thank J.~P.~Garrahan and Y.~Kafri for insightful discussions. All authors are associated with Centre for Quantum Engineering at Aalto University. K.~B.~acknowledges financial support from Academy of Finland (Contract No.~296073). The work was supported by Academy of Finland (Project No.~308515).
\end{acknowledgments}

\appendix

\section{Product expansion \label{app:appB}}
\begin{figure}
	\centering
	\includegraphics[width=0.9\columnwidth]{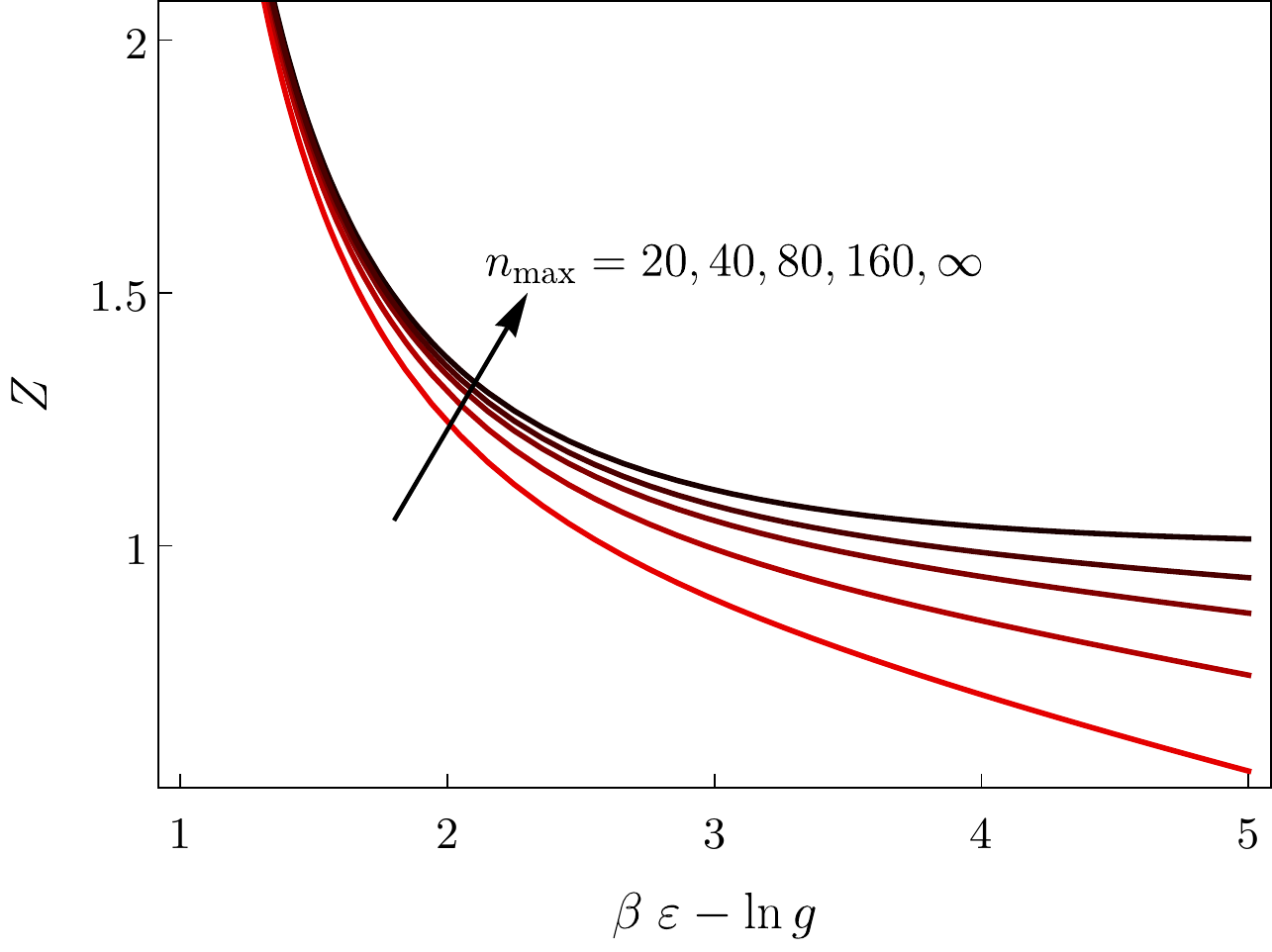}
	\caption{\label{fig:prod_exp} Product expansion of the partition function. The dashed lines show Eq.~\eqref{ZipperProdExp} for $N=20$. We have truncated~$n$ from $- n_{{{\rm max}}}$ to $+n_{{{\rm max}}}$. With increasing values of $n_{{{\rm max}}}$, the product expansion converges towards the exact partition function for the zipper model shown with a black line.}
\end{figure}

The product expansion of an entire function $f(z)$ around $z=0$ has the general form\cite{arfken2012}
\begin{equation}\label{GenProdExp}
f(z) = f(0)e ^{z f'(0)/f(0)}
\prod_n \left(1-\frac{z}{z_n}\right) e^{z/z_n},
\end{equation}
where the prime indicates the first derivative with respect to $z$ and the product runs over all complex zeros $z_n$ of $f(z)$. For the partition function in Eq.~(\ref{eq:PartFunct}), the product expansion explicitely reads
\begin{equation}
Z= \frac{1-g^{N+1}}{1-g}e^{\beta\varepsilon A_N}
\prod_{n=-\infty}^\infty\prod_{m=1}^N
\left(1-\frac{\beta\varepsilon}{a_{nm}}\right)
e^{\beta\varepsilon/a_{nm}},
\label{ZipperProdExp}
\end{equation}
where we have introduced
\begin{equation}
A_N= (N+1)\frac{g^{N+1}}{1-g^{N+1}}-\frac{g}{1-g}
\end{equation}
and
\begin{equation}
a_{nm}= \ln g -2\pi i\left(\frac{m}{N+1}+n\right).
\end{equation}
In Fig.~\ref{fig:prod_exp} we compare the exact expression for the partition function with the production expansion Eq.~\eqref{ZipperProdExp}. The production expansion convergences towards the exact result as an increasing number of zeros are included.

\section{Probability density \label{app:appA}}

We consider a system with $N$ sites in the canonical ensemble with energies $E_n$ and corresponding degeneracy factors $g_n$. The partition function can be written as
\begin{equation}
Z(\beta)=\sum_{n=1}^{\Omega(N)} g_n e^{-\beta E_{n}},
\end{equation}
where $\beta=1/(k_BT)$ is the inverse temperature and $\Omega(N)$ is the number of possible energies. For large systems, $N \ \gg 1$,  the partition sum can be approximated as
\begin{equation}
\begin{split}
Z(\beta)&= N \sum_{n=1}^{\Omega(N)} \frac{g_n}{N}  \ e^{-\beta N (E_{n}/N)}\\
 &\simeq N \int_{0}^{u_\mathrm{max}} du \ g(u) \ e^{-\beta N u},
\end{split}
\end{equation}
where $u_\mathrm{max}$ is the maximal energy per site and $g(u)$ is the density of states. Since the Boltzmann factor decays rapidly for large $N$, we can write the partition sum as
\begin{equation}
Z(\beta) \simeq N \int_{0}^{\infty} du \ g(u) \ e^{-\beta N u}
\end{equation}
by extending the upper limit of the integral to infinity.

From the partition sum, we can identify the probability density of finding the system with energy $u$ per site as
\begin{equation}
P(u)=\frac{N g(u) \ e^{-\beta N u}}{Z(\beta)}.
\end{equation}
It is now convenient to consider the ratio
\begin{equation}
X(\gamma) = \frac{Z(\beta+\gamma)}{Z(\beta)}\simeq \int_{0}^{\infty} du P(u) e^{-\gamma N u}
\end{equation}
of the partition function at inverse temperatures $\beta+\gamma$ and $\beta$. Specifically, we see that $X(\gamma)$ can be considered as the Laplace transform of the probability density $P(u)$.\cite{Olver1974} Thus, through an inverse Laplace transformation, we find
\begin{equation}
P(u)\simeq \frac{N}{2 \pi i} \int_{c - i \infty}^{c + i \infty} d\gamma   X(\gamma) e^{\gamma N u}
\end{equation}
where the real constant $c$ has to be chosen such that the integral $\int_{0}^{\infty} du  P(u) e^{-c u}$ is finite. Since $\int_{0}^{\infty} du P(u)=1$, we are free to set $c=0$. Finally, by a change of variable we find the expression used in the main text
\begin{equation}
P(u)\simeq \frac{N}{2 \pi} \int_{-\infty}^{+\infty} d\gamma X(i \gamma)e^{i \gamma N u}.
\end{equation}

\end{document}